\documentclass[a4paper]{article}
\usepackage[utf8]{inputenc} 
\usepackage[T1]{fontenc}
\usepackage{todonotes}
\usepackage{graphicx}
\usepackage{float}
\usepackage{url}

\title{OpenStack and Google Cloud performance comparison in Infrastructure as a Service model}

\author{Michał Łątkowski\footnote{ORCID 0000-0002-2663-7312} and Robert Nowak\footnote{ORCID 0000-0001-7248-6888}\footnote{Correspondence: robert.nowak@pw.edu.pl}}

\begin{document}

\maketitle

\begin{abstract}
  Cloud computing is becoming common, and the choice of proper infrastructure is essential. One of main issues is choosing between private and public clound, between commercial and non-commercial solutions. This paper aims to compare the parameters of OpenStack and Google Cloud systems. Both systems deliver a computing cloud service, enabling the user to use the infrastructure as a service (IaaS) model. We developed the pipeline using the Python programming language and its libraries, which enable communication with the aforementioned clouds. We measured various parameters of instances and task execution: instance launch and deletion times, and their dependence on the number of launched instances. Moreover, we used benchmark algorithms to check the instance performance. We analysed the results and the factors that contributed to them and provided conclusions, recommendations, and suggestions for further research based on the gathered data.

\end{abstract}



\section{Introduction}

Cloud computing is a model of universal, convenient, network-based access at the user's request to a shared pool of configurable computing resources (networks, servers, applications, services) that can be quickly made available with minimal interaction by the service provider \cite{mell2011nist}.
The constantly growing popularity of all clouds suggests the need for extensive, objective research, analysis, and comparisons of individual cloud solutions.

Infrastructure as a Service (IaaS) is a model in which the user has access to networks, disk space, processing nodes, etc. The user can run any operating systems and applications and, to a limited extent, manage network aspects of the resources offered. However, the user still does not physically control the infrastructure \cite{mell2011nist}.

In our work, we consider two cloud technologies - Google Cloud Platform (GCP)\cite{challita2018precise, bisong2019overview} and OpenStack\cite{rosado2014overview, openstackintroduction2021}. The reason for choosing these clouds was, respectively, OpenStack's unflagging position as the leading open-source cloud computing software and the growing popularity of Google Cloud Platform. In the \cite{flexera2021} report, Google Cloud Platform, compared to Microsoft Azure and Amazon Web Services, achieved the greatest increase in its use by large companies. An additional motivation is that while OpenStack is a tool widely researched in the literature, few scientific articles analyse Google Cloud in the context of IaaS systems (compared to, for example, articles about Amazon Web Services).

OpenStack's position as one of the leading technologies in the private cloud market has been reinforced by numerous studies. Much of the research (for example \cite{yadav2013comparative}) compared the features offered by individual private clouds as well as their architectures. However, we would like to focus more on the performance aspect.

In this article, we compare the parameters of OpenStack and GCP.
The motivation of this paper is:
\begin{itemize}
\item An explicit comparison between those 2 technologies has not been carried out before.
\item The available literature depicts the performance before 2017, and both technologies have been changed multiple times throughout 5 years.
\item The conclusions are very significant for research teams at universities. Our findings support those researchers trying to decide whether to buy physical infrastructure and maintain a private cloud or use a public cloud, paying for its' use.
\item  Some characteristics we analyzed have not been studied in other papers. What is more, we tried to eliminate the influence of other factors while studying each element.
\end{itemize}

We developed the pipeline using the Python programming language as depicted in Sec.~\ref{sec:methods}.
The measured parameters of instances and task execution are depicted in Sec.~\ref{sec:results}. We provided a wide range of parameters and used the same tools and benchmarks for both cloud systems. We conclude our research in Sec.~\ref{sec:conclusion}, Sec.~\ref{sec:discussion} depicts comments and further works.

Presented research also verifies the extent to which other papers' conclusions have become outdated since each research we found on the aforementioned clouds was published at least four years ago.

\section{Literature review}

The article \cite{mao2012performance} addresses the issue of instance startup time. The authors examine AWS, Azure, and Rackspace clouds. OpenStack is not directly mentioned in this work, but Rackspace was considered, and due to the fact that Rackspace was built on OpenStack, the results might apply to OpenStack. The authors share many interesting conclusions.
\begin{itemize}
\item Particular attention should be paid to the information provided by cloud providers. For example, the fact that an instance is in the running state does not necessarily mean it is ready to be used.
\item Instance startup time does not depend on the time of day.
\item Instances running MS Windows in Rackspace take significantly longer to start up (9 times longer in the study) than those running Linux.
\item Instance startup times are positively correlated with the system image size.
\end{itemize}
We explored all these aspects for both clouds (OpenStack and GCP) in this paper.

The paper \cite{vogel2016private} strictly compares three private clouds: OpenNebula, CloudStack and OpenStack. Although the authors emphasize that clouds can be compared in different ways, the most important conclusions in this work are related to efficiency. Of the three, OpenStack is the most stable. The results are surprising as all three private clouds showed similar performance results to those obtained when running tests in a native environment, i.e. without virtualization. The lack of differences stems from choosing peculiar criteria, more distant than those used in the other studies and those used in this study.

Interesting correlations can be found in the \cite{6899177} study comparing OpenStack and CloudStack, such as a positive correlation between instance startup and deletion times and disk size. The conclusion is as follows -- OpenStack is significantly better in terms of startup and deletion times of instances and the performance presented in benchmarks.
These results are confirmed in another comparison of the same platforms~\cite{mullerikkal2015comparative}. According to the authors, OpenStack rewards the user for significant difficulties in the initial configuration with better benchmark results and overall stability and performance. The authors used the UnixBench tool~\cite{unixbench2021}. OpenStack achieves a significant advantage in almost every test that is part of this tool.

The \cite{steinmetz2012cloud} demonstrates the advantage of OpenStack over Eucalyptus in performance tests (also carried out using UnixBench). An important conclusion is the lack of a negative correlation between the number of simultaneously launched instances and the average time of launching one instance. The conclusion suggests that OpenStack was not running instances concurrently at the time of the study. We analyse this aspect in more depth in the following sections.

The article~\cite{bootcomputeengine2021} presents the analysis of the instance startup times on GCP.
The startup time is from issuing the command to create an instance until the instance is ready to go into standby (equivalent to the fact that the user can connect to the instance, for example, via SSH). The author distinguishes three phases in this process: query processing, resource allocation, and instance launch. The consequence of the analysis presented in~\cite{bootcomputeengine2021} is the following conclusion - among the three mentioned phases, the first two are relatively constant - that is, the user has no practical influence on these phases. The greatest fluctuations occur at the stage of starting the instance.
The number of CPU cores and RAM does not matter in terms of the startup time of the instance. The boot time depends on the system image type - with the emphasis that Linux images behave similarly. The most significant differences appear when we compare Windows images with Linux images.

In the article \cite{leitner2016patterns}, the authors formulate 15 different hypotheses regarding the performance of IaaS systems and the reasons for the differences in performance between them. There are also two important concepts not previously mentioned in this paper: hardware heterogeneity and noisy neighbour effect.
Hardware heterogeneity means using different types of resources, for example, processors with different architectures, to support the same type of instance. Some studies, e.g. \cite{181159}, indicate that it may influence the observed results.
The noisy neighbour effect is a deterioration in the performance of the instance in use due to sharing the physical infrastructure with some other instances currently taking up the resources. Usually, the processor cores are not shared, but there are differences in reading and writing from disk or network bandwidth.

The geographic location of the instance does matter, while temporal factors (such as time of day, day of the week) do not have a significant influence. However, it should be emphasized here that the differences between regions do not transfer between different clouds, i.e. a region that offers better performance than the other for one technology may be worse when using another cloud. In the studied clouds, the European region is usually preferable in terms of performance to the American region.

The article \cite{ahuja2020multi} compares GCP and AWS in terms of various factors, such as compute, memory read/write, disk I/O and other factors not analyzed in this paper (such as pricing, elasticity or network bandwidth). The authors, by running series of distributed benchmarks on Amazon Web Services and Google Cloud Platform. The conclusions were: AWS excelled in memory bandwidth, but Google Compute Engine surpassed Amazon's EC2 in terms of read and write throughput. The CPU performance did not differ significantly.

We also studied reports from external companies that compare the performance of instances offered by individual vendors. The Cockroach Labs report from 2021 \cite{cockroach2021} compares AWS, Azure, and GCP. Using a series of benchmarks, the authors of this report reach the following conclusions:
(1) if we are measuring the performance of single-core processors, GCP wins over others; (2) AWS scales better than GCP in the case of multi-core processors, as well as in the case of database transactions; (3) Azure is the best at I/O operations.
The final winner of the above comparison in 2021 was GCP, which suggests there is a need for further research on this technology.

\section{Materials and Methods}
\label{sec:methods}

\begin{figure}[!htb]
    \centering \includegraphics[width=1\linewidth]{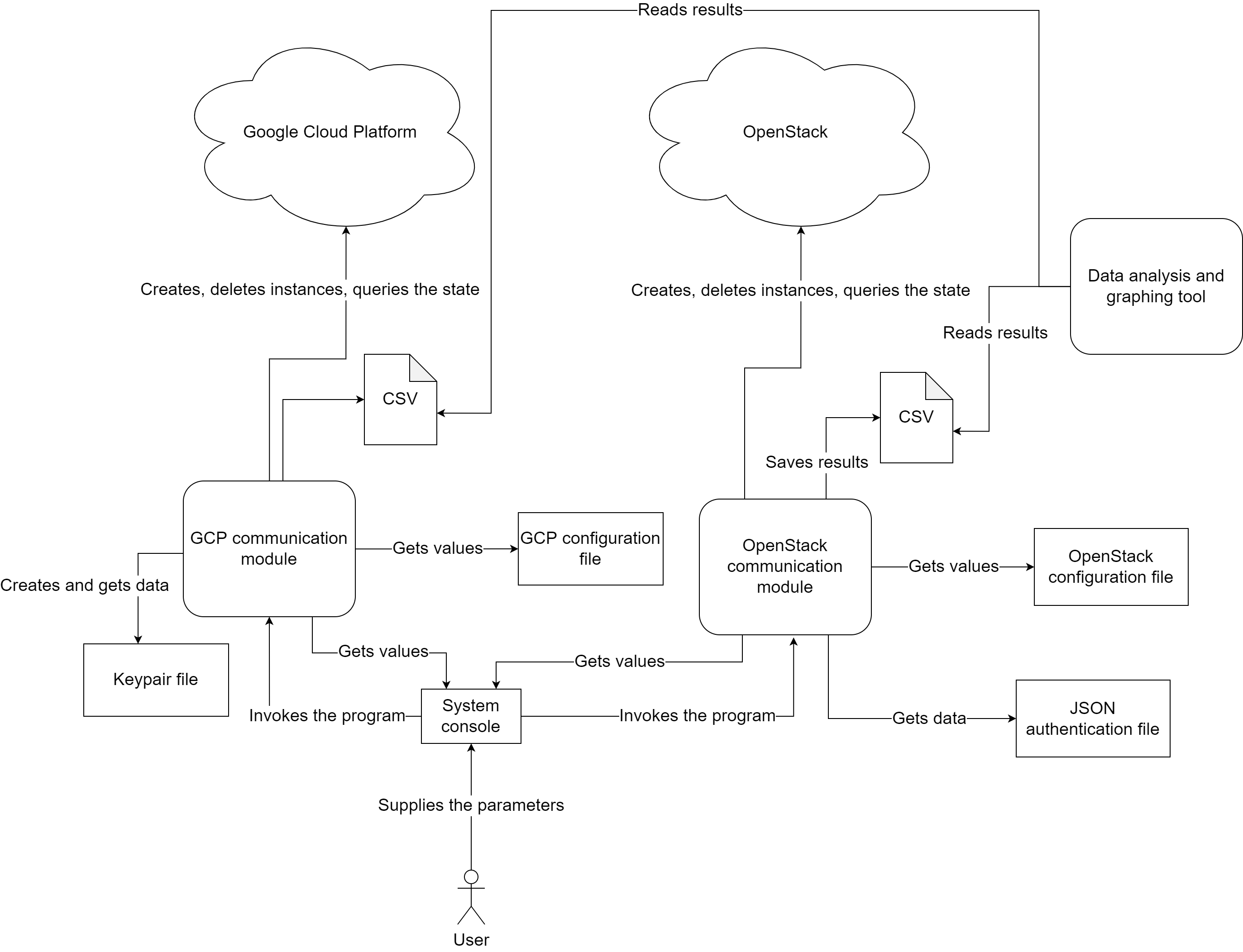}
    \caption{Diagram of interaction between the user and application modules, direction of the arrows -- from the caller to the called.}
    \label{fig:schemat_modulowy_aplikacji}
\end{figure}

The pipeline we created to carry out the experiments consists of 3 tools: application for GCP measurements, application for OpenStack, and application for results analysis. All applications were made in Python.
The GCP part uses the \textit{googleapiclient.discovery} library to create an object that connects to the GCP cloud. We developed modules to create and delete instances, perform various measurements, and store results in the database.
The OpenStack application uses \textit{openstacksdk} library.
The results analysis application uses a \textit{Jupyter Notebook}\cite{jupyter} and its visualization part with our extensions.

Interactions between the user and the pipeline are presented in the Fig.~\ref{fig:schemat_modulowy_aplikacji}.
Our applications are used in a similar way as many Unix systems tools, where the modification of parameters is performed by setting appropriate flags.
By modifying the parameters, the user can replicate all the measurements described in Section~\ref{sec:results} using the appropriate shell commands.
The detailed user instruction including a list of flags that were used to carry out the experiment is in Appendix~\ref{app:parameters}.
The performed measurements are saved to CSV files.
Each record includes startup time, deletion time, and timestamp.

\begin{figure}[!htb]
    \centering \includegraphics[width=.5\linewidth]{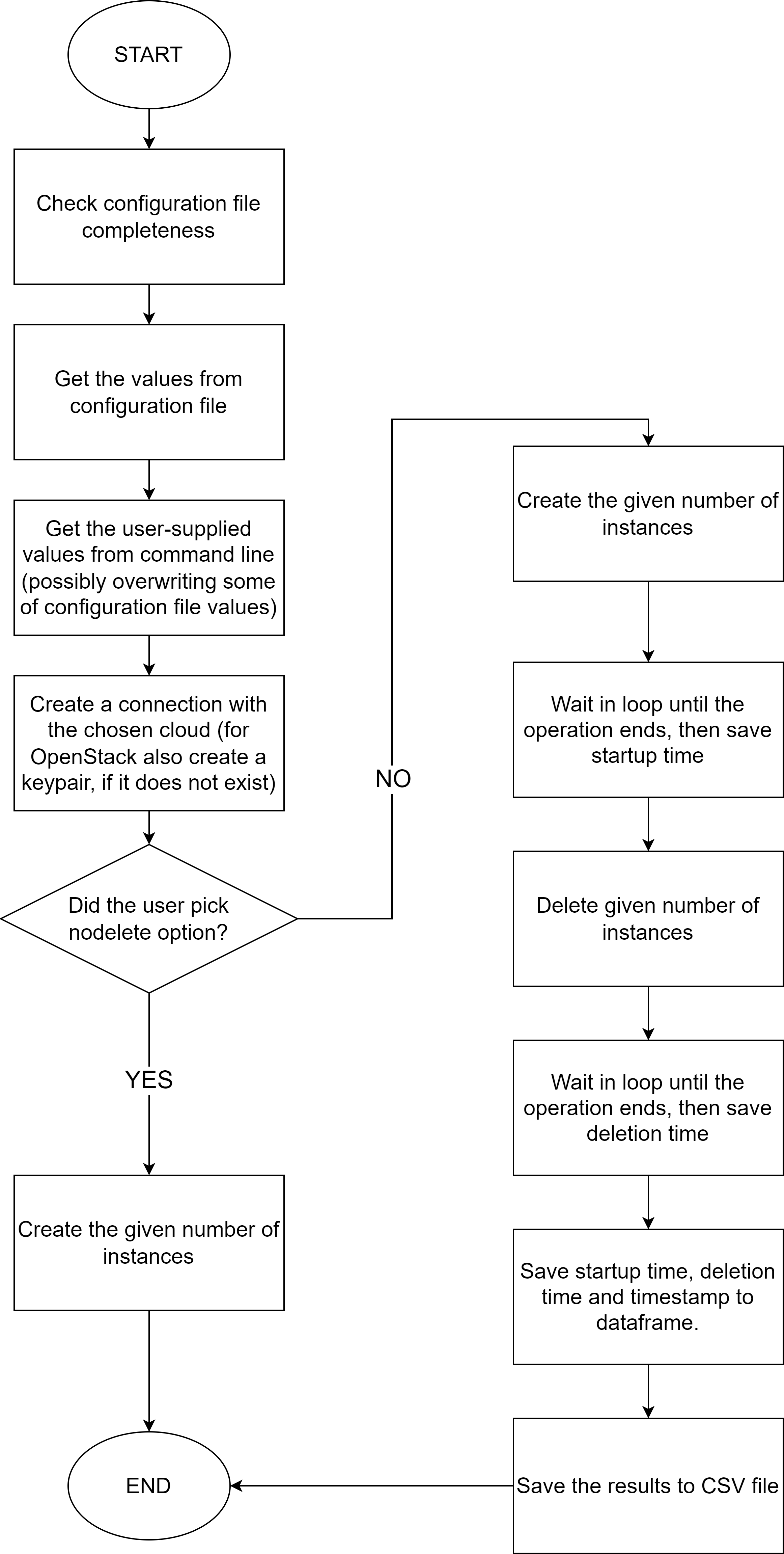}
    \caption{Application workflow diagram.}
    \label{fig:schemat_dzialania_aplikacji}
\end{figure}

GCP measurements application and OpenStack measurements application algorithm is similar and is shown in Fig.~\ref{fig:schemat_dzialania_aplikacji}.

\section{Results}
\label{sec:results}

Our experiments reduce random factors influencing the measurements by repetition. What is more, we always try to keep the other parameters (apart from the measured one) exactly the same.
Each experiment, apart from the one that examines the influence of the zone (and region), is performed in the \textit{us-central1-a} zone in the case of Google Cloud Platform.
Each of the instance startup time measurements was repeated 100 times. The measurements of benchmarks were repeated 10 times.
Such an approach reduces most of the factors, such as the influence of the state of the provider's services or noisy neighbour problem. 

For our experiments, we defined several instances depicted in Tab.~\ref{tab:instances}.

\begin{table}[!htb]
  \center
  \begin{tabular}{l|r|r|r|p{0.5\linewidth}}
    name & vCPU & RAM & disk & comments \\ \hline
    f1-micro     & 1 $\times \*\* $ &  614 MB & 10 GB & a shared-core instance, GCP instance \\
    n1-standard-1  & 1 $\times$ &  3.75 GB & 10 GB & contrary to OpenStack, GCP instances are not tied with any particular disk size. However, in this paper, GCP instances use 10 GB disks \\
    n1-standard-8  & 8 $\times$ & 30 GB & 10 GB & GCP instance \\
    custom  & 8 $\times$ & 16 GB & 16 GB & GCP instance \\
    os1 & 1 $\times$ & 1 GB & 5GB & OpenStack instance \\
    os2 & 1 $\times$ & 1 GB & 10 GB & OpenStack instance \\
    os3 & 4 $\times$ & 4 GB & 10 GB & OpenStack instance \\
    os4 & 4 $\times$ & 4 GB & 32 GB & OpenStack instance \\
    os5 & 8 $\times$ & 16 GB & 10 GB  & OpenStack instance\\
    os6 & 8 $\times$ & 16 GB & 64 GB & OpenStack instance \\
    os7 & 16 $\times$ & 64 GB & 64 GB & OpenStack instance \\
  \end{tabular}
  \caption{Instances used in numerical experiments. The description includes the number of virtual CPUs, RAM memory size, and disk space. If not mentioned otherwise, the instance has the Ubuntu 20.04 system.
  \label{tab:instances}
  }
\end{table}

\subsection{Dependence of instance launch times on the time of day}

\begin{figure}[!htb]
  \centering
  \includegraphics[width=0.7\textwidth]{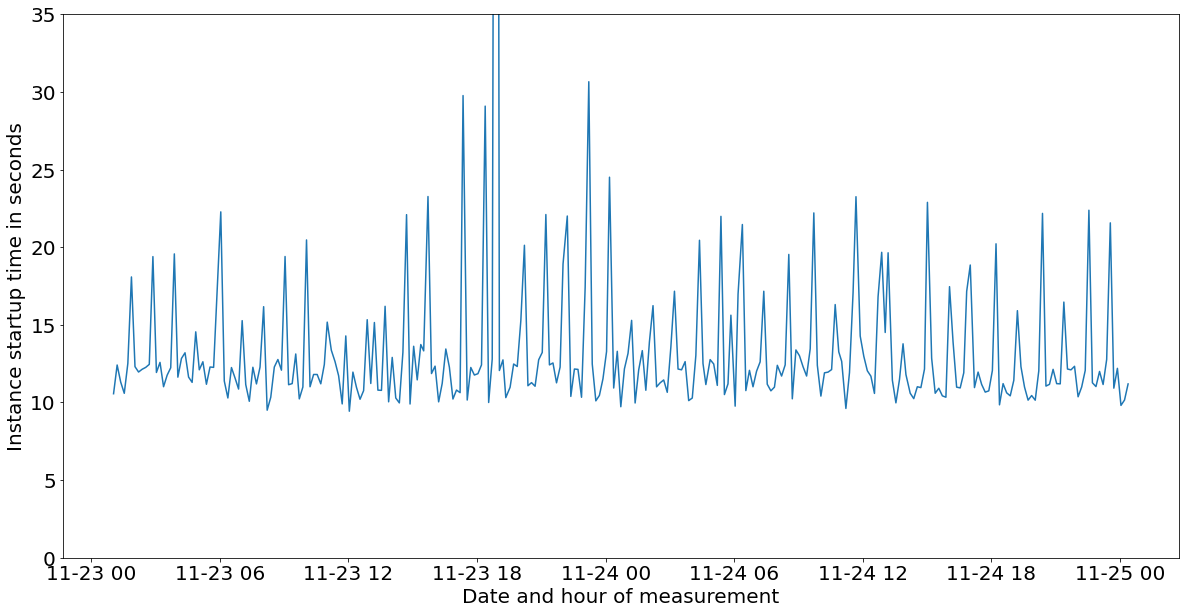}

  \includegraphics[width=0.7\textwidth]{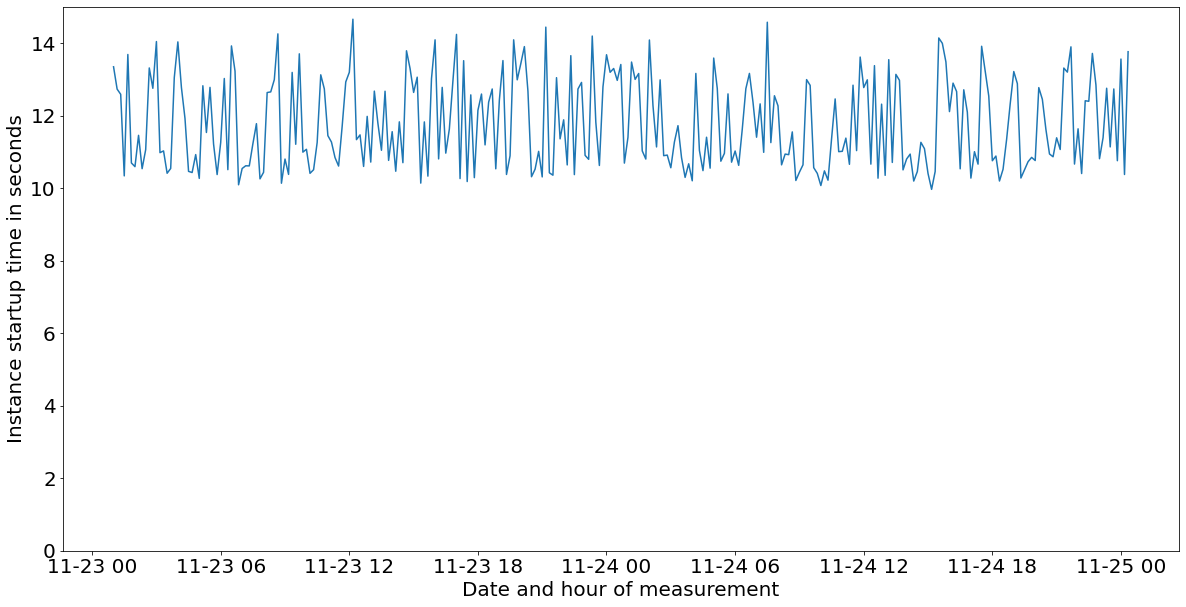}

  \caption{Dependence of instance launch times on the time of day, GCP with \emph{n1-standard-1} on the left, OpenStack with \emph{os1} on the right.}
  \label{fig:pora_dnia_czas}

\end{figure}

Both in GCP and OpenStack, the time of the day has no effect on instance startup times, as depicted in~Fig.~\ref{fig:pora_dnia_czas}.
The difference between days of the week is also not important.
The reason for this could be, in both cases, infrastructure redundancy. Both clouds had enough resources at any point in time during the analysis period.
However, in the case of private clouds, which often do not have service level agreements,
one should take into account the hardware capabilities of the infrastructure, as well as whether intensive calculations are performed during measurements.

Our results confirm the conclusions of the article \cite{leitner2016patterns}. Considering the above experiment results, in further experiments, the times of the day and days of the week are treated as negligible factors in OpenStack and GCP.




\subsection{Dependence of startup times on the region in GCP}

We compared the times of lauching the instance when it is placed in \textit{us-central1-a} zone to \textit{europe-central2-a}.
In both cases, an instance of type \textit{n1-standard-1} was used.

\begin{figure}[!htb]
    \centering \includegraphics[width=1\linewidth]{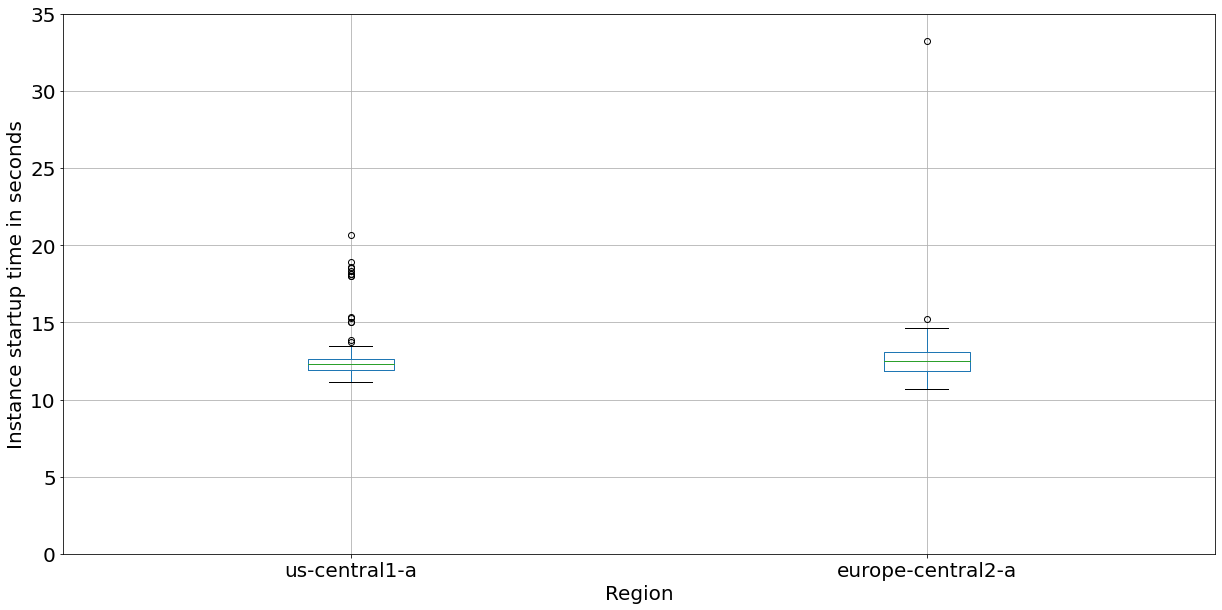}
    \caption{GCP - dependence of instance launch times on the region}
    \label{fig:gcp_region_czas_box}
\end{figure}

The Fig.~\ref{fig:gcp_region_czas_box} shows the results.
Not only do instances in the \textit{europe-central2-a} zone take less time to start up on average than those in \textit{us-central1-a} ($12.7$ versus $13.5$),
but they also have different variances (9.7 in the US and 5.0 in Europe).
Our results confirmed conclusions from \cite{leitner2016patterns}.

A single measurement in the European zone, which differs from the average, was made at 23:07:51, on December 17, 2021. Most likely, it was
due to network problems or a temporary delay in service delivery from GCP, as other measurements made that day are close to average.

In our opinion, the difference in launching time is the data centre load in a given zone. The zone in Warsaw is relatively new at the time of writing, it was opened in 2021, and the hardware in this zone is not under significant load. In turn, the studied American region is older. Its age is already around 5 years, so its load is bigger. Secondly, the prices of instances in both zones are different, which affects their popularity.

The results depict that the region is an important factor. We eliminate this factor from other experiments,
as the zone is the the \emph{us-central1-a} for next experiments.
OpenStack's geographical comparison was not performed because we used the Warsaw University of Technology private cloud.

\subsection{Dependence of instance launch times on machine type}
\label{sec:machine-type}

In the experiment below, we will examine the effect of the machine type, i.e. the number of virtual processors, RAM size, etc.
on the instance startup times.
In GCP we compare: \emph{f1-micro}, \emph{n1-standard-1}, and \emph{n1-standard-8} (from~Tab.~\ref{tab:instances}),
In OpenStack we compare: \textit{os2}, \textit{os3}, and \textit{os5}.

\begin{figure}[!htb]
  \centering

  \includegraphics[width=0.7 \textwidth]{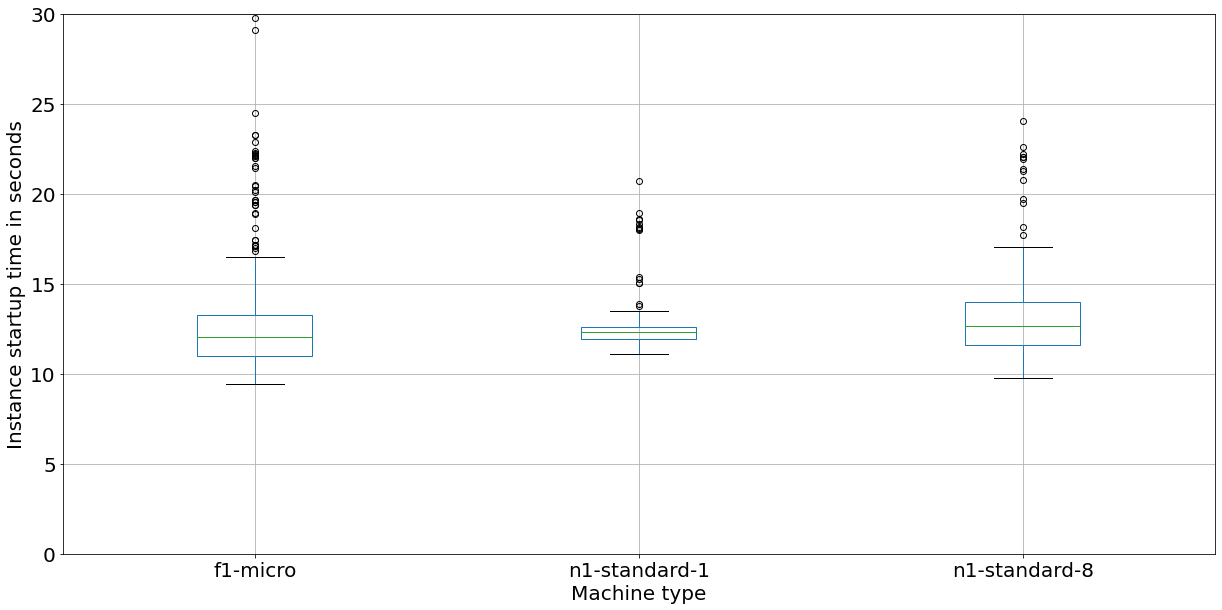}

  \includegraphics[width=0.7 \textwidth]{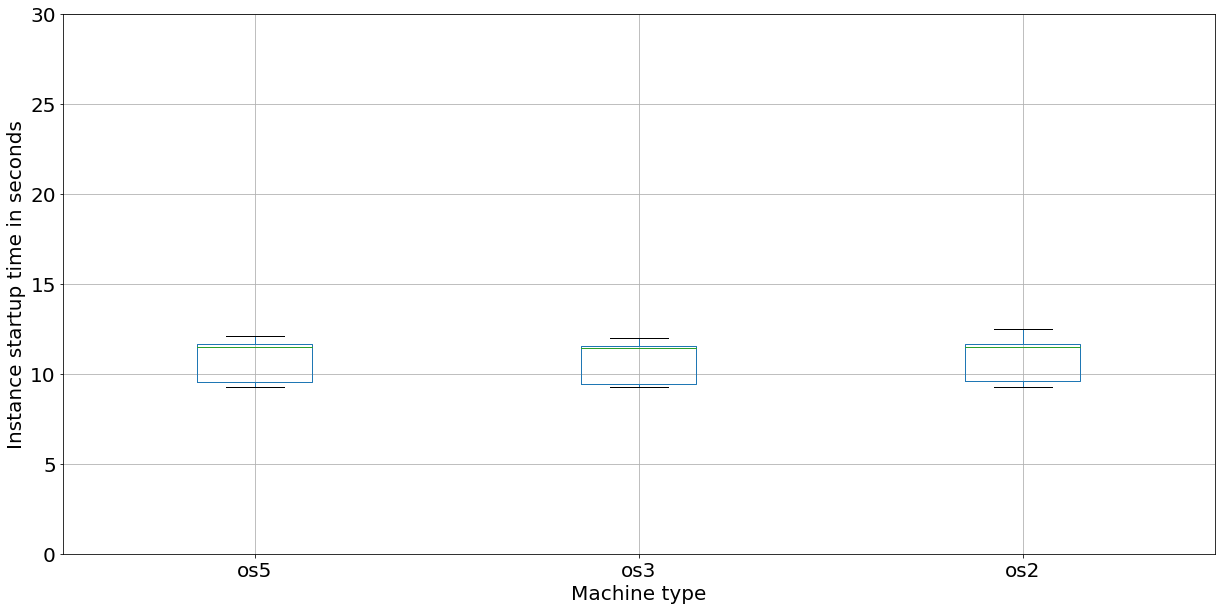}

  \caption{Dependence of instance launch times on machine type, GCP on the left, OpenStack on the right.}
  \label{fig:typ-maszyny-czas}
\end{figure}

As can be seen in~Fig.~\ref{fig:typ-maszyny-czas}, we have the average launch times (in seconds): $13.6$ (n1-standard-8), $12.9$ (n1-standard-1), $13.5$ (f1-micro), $10.8$ (os2), $10.7$ (os3), $10.9$ (os5).
There is no correlation between machine type and startup times.
This means that the bigger machines startup just as quickly as the smaller ones, contrary to intuition.
Our results confirm the findings from the study for GCP \cite{googlemachinetypes2021}.
In the case of OpenStack, our findings are in contradiction to the conclusions of \cite{6899177}. However, the authors themselves admitted that the correlations were not strong.
In the article \cite{mao2012performance}, mere observation of the graphs clearly shows that instance launch times on Rackspace did depend on machine type, while such correlation was not apparent in AWS.
It's worth noting, however, that Rackspace may or may not display similar behaviour to OpenStack -- these are not the same technologies.
OpenStack is still under development and $20$ new editions have appeared since 2012.
The outdating of the above research becomes even more likely when looking at the instance startup times. The instance, which the authors named \emph{type 1} (256 MB RAM and 10 GB disk) started on average in 35 seconds. In our experiments, a bigger instance starts in about $10$ seconds.

\subsection {Dependence of startup times on the disk size}

The number of virtual processors and RAM does not influence the instance startup times, as depicted in Section~\ref{sec:machine-type}.
To compare the startup times dependence on disk size, we used three instances in GCP, all of type \textit{n1standard}, varying in disk sizes: 10GB, 50GB, and 500GB.
We also present the results of a similar experiment in the OpenStack cloud,
using \emph{os1}, \emph{os4}, \emph{os6} and \emph{os7}.

\begin{figure}[!htb]
  \centering

  \includegraphics[width=0.7 \textwidth]{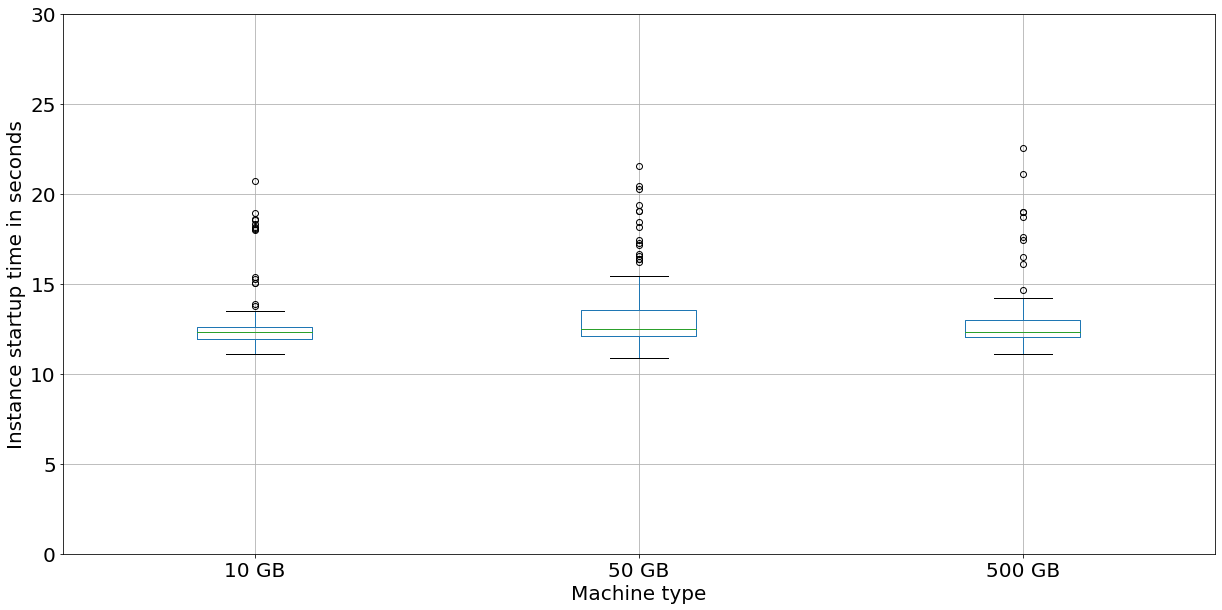}
  \hfill
  \includegraphics[width=0.7 \textwidth]{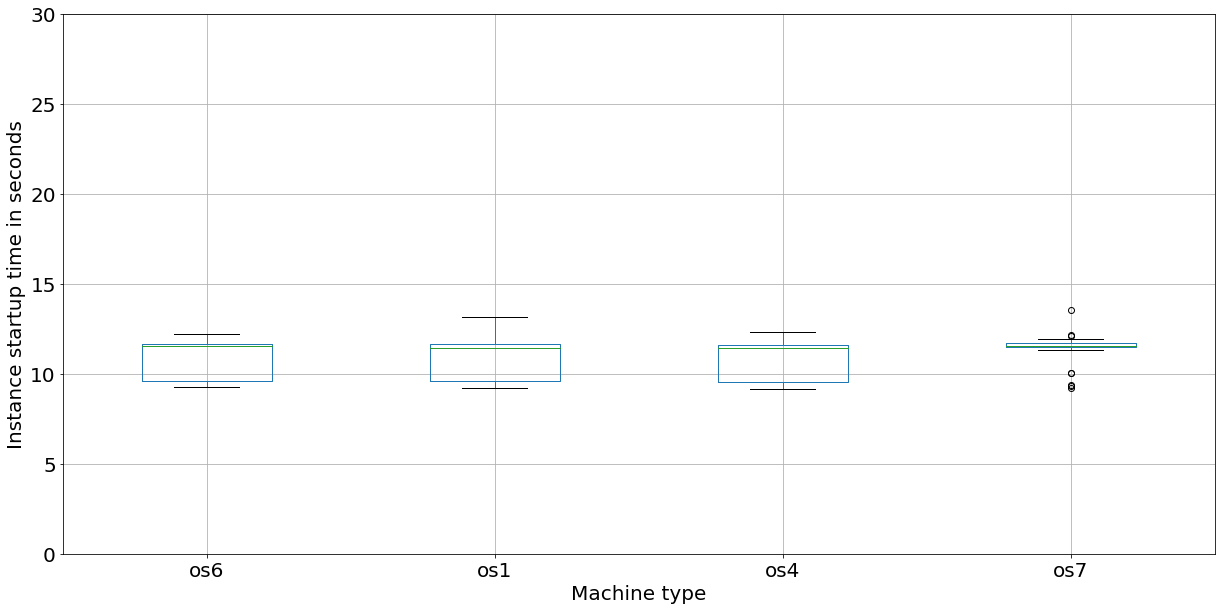}

  \caption{Dependence of startup times on the instance disk size, GCP on the left with \emph{n1-standard} instance, OpenStack on the right}
  \label{fig:rozmiar_dysku_box}
\end{figure}

By analysing Fig~\ref{fig:rozmiar_dysku_box}, we get a clear conclusion -- the disk size does not seem to affect while the instance is running, the average times (in seconds) are:      $12.9$ (10GB), $13.4$ (50GB), $12.9$ (500GB), $11.5$ (os1, 5GB),  $10.9$ (os2, 10GB),  $10.7$ (os5, 10GB),  $10.8$ (os6, 64GB).
Combining machine type and disk size also does not result in significantly different values.

\subsection{Dependence of instance startup time on the system}

An important element that should be taken into account when setting the parameters of an instance is the operating system.
In GCP we used \textit{n1-standard-1} machines, with Ubuntu, Debian, and Clear Linux systems, where Clear Linux is a system with very small size and requirements.
In OpenStack, we used \textit{os1} machines, and we decided to compare two systems -- Ubuntu and Cirros,
where Cirros is a lightweight system designed for the needs of OpenStack, mainly for testing purposes.

\begin{figure}[!htb]
  \centering

  \includegraphics[width=0.7 \textwidth]{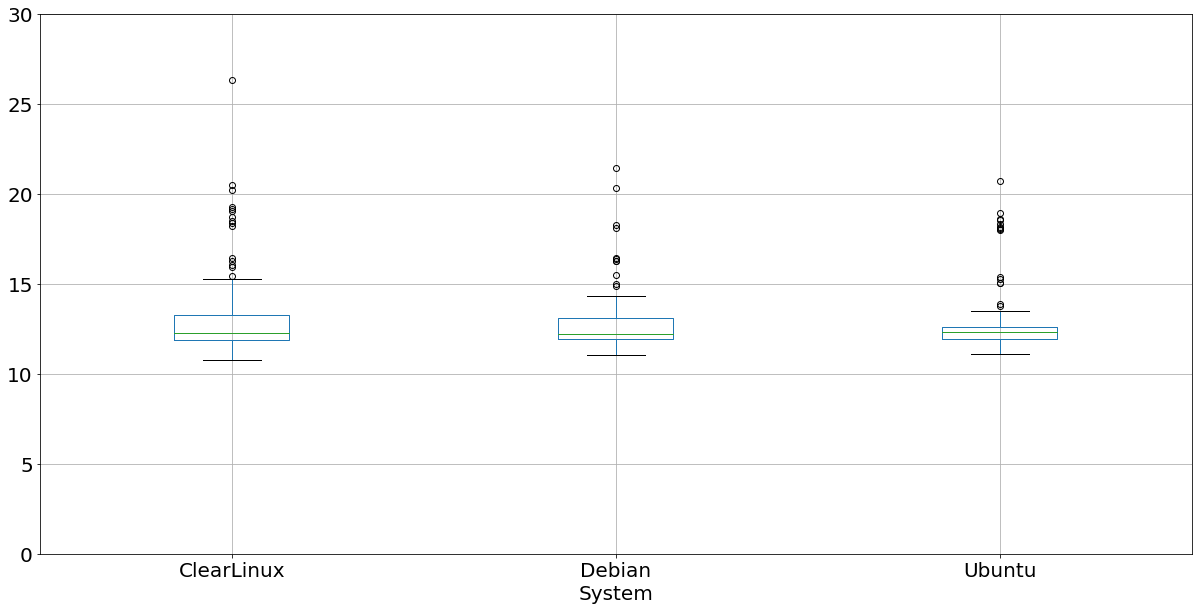}

  \includegraphics[width=0.7 \textwidth]{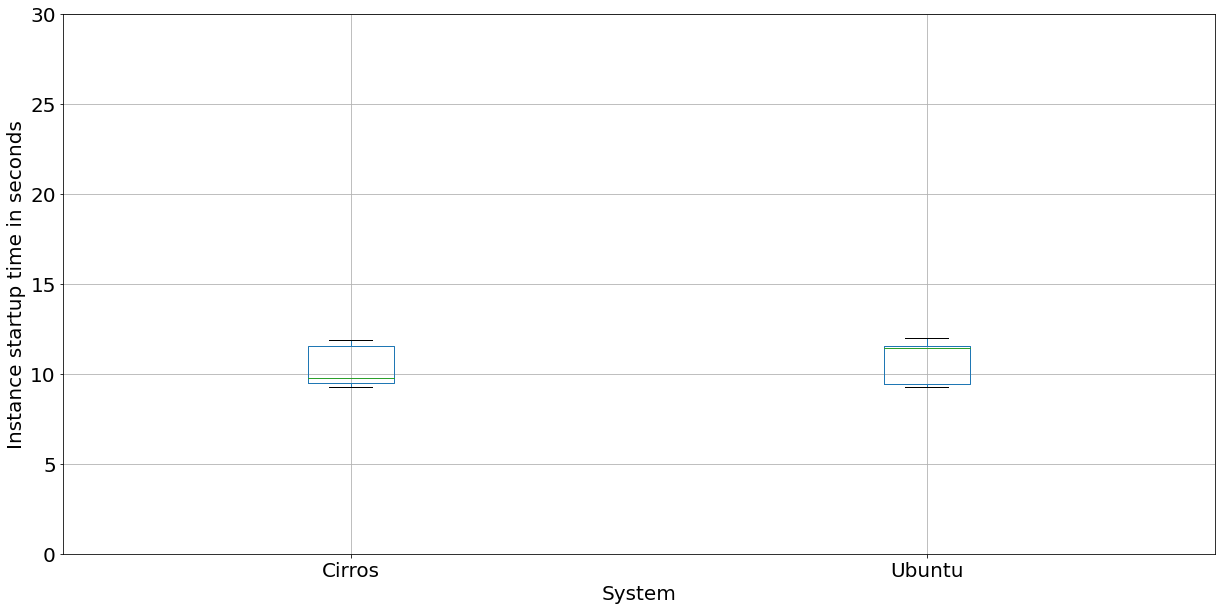}

  \caption{OpenStack - Dependence of startup times on the system, GCP on the left, OpenStack on the right.}
  \label{fig:system_czas_box}

\end{figure}

\begin{table}[!htb] \centering
\begin{tabular} {l|l|r}
    Cloud &  System & Average [s]  \\ \hline
    GCP   & Ubuntu & 12.9 \\
    GCP   & Debian & 13.1 \\
    GCP   & Clear Linux & 13.2 \\ \hline
    OpenStack & Ubuntu & 10.7 \\
    OpenStack & Cirros & 10.5 \\
\end{tabular}
\caption{Dependence of startup times on the operating system.}
\label{tab:os-systems}

\end{table}

The differences in the operating system from the Linux family installed on the instance are negligible and do not correlate with the size of the operating system.
The results are depicted in~Fig.~\ref{fig:system_czas_box} and in~Tab~\ref{tab:os-systems}.
Both clouds work similarly. They store ready-made images, and starting up systems does not take much time, regardless of their size.
The average time for launching a Linux machine is $12.9$ s.

\begin{figure}[!htb]
  \centering
  \includegraphics[width=0.7\linewidth]{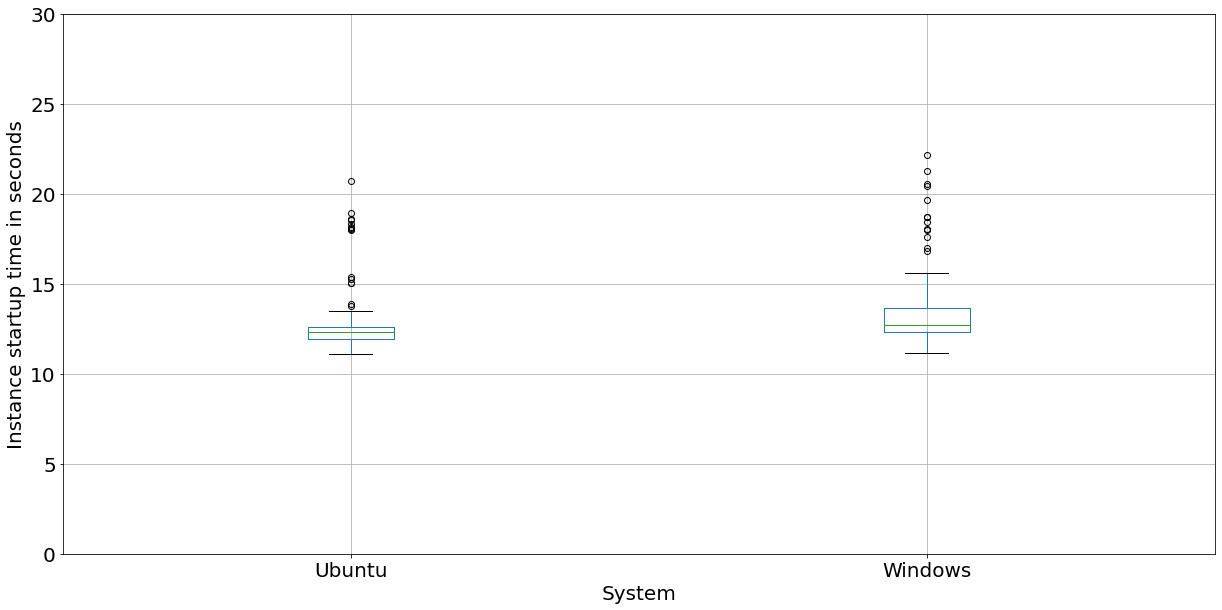}
    \caption{GCP - difference between Windows and Linux in terms of instance startup time on GCP.}
    \label{fig:gcp_windows_czas_box}
\end{figure}

We also investigate the instances with MS Windows operating system. The results are depicted in~Fig.~\ref{fig:gcp_windows_czas_box}.
We present only GCP with \emph{n1-standard-1}. The average time launching for MS Windows instance is $13.6$ seconds.



\subsection{Dependence of startup times on the number of instances in GCP}

GCP \emph{bulkInsert} function allows the user to start up multiple instances at once. In this study, we check whether starting up multiple instances at once has a significant impact on the speed of their launch. We are using \textit{n1-standard-1} instance.

\begin{figure}[!htb]
  \centering
  \includegraphics[width=0.7\linewidth]{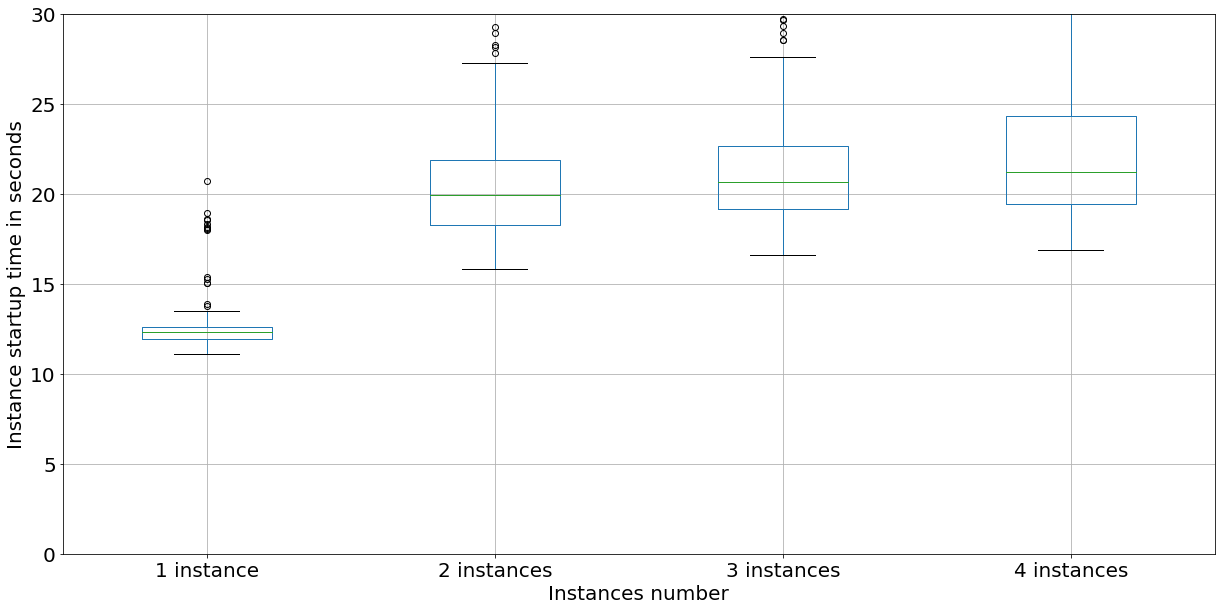}
    \caption{GCP - dependence of startup time on the number of instances}
    \label{fig:gcp_wiele_instancji_czas_box}
\end{figure}

The results shown in Fig.~\ref{fig:gcp_wiele_instancji_czas_box} are as expected. The number of instances significantly changed the startup time of an individual instance. For a single instance, we got on average $12.9$ seconds of startup time; for two instances: $20.7$ s; three instances: $21.6$ s; and four instances $21.9$ s.
The main reason for such results is, in our opinion, the fact that the speed of starting the instance groups is as high as the slowest virtual machine's speed.

\subsection{Dependence of startup times on preemptibility in GCP}

The preemptible instances should theoretically be a good choice for users who need instances for a short period and do not care about system stability.
It is worth noting that GCP does not guarantee the availability of these instances or any level of quality of the services provided in relation to them.
Moreover, the company reserves the right to stop the instance at any time (and after 24 hours they stop automatically).
The experiment depicted in this section examines whether preemptible instances have the same launch times as their non-preemptible counterparts.
For this we are using \textit{n1-standard-1} instance with Clear Linux.

\begin{figure}[!htb]
  \centering
  \includegraphics[width=0.7 \linewidth]{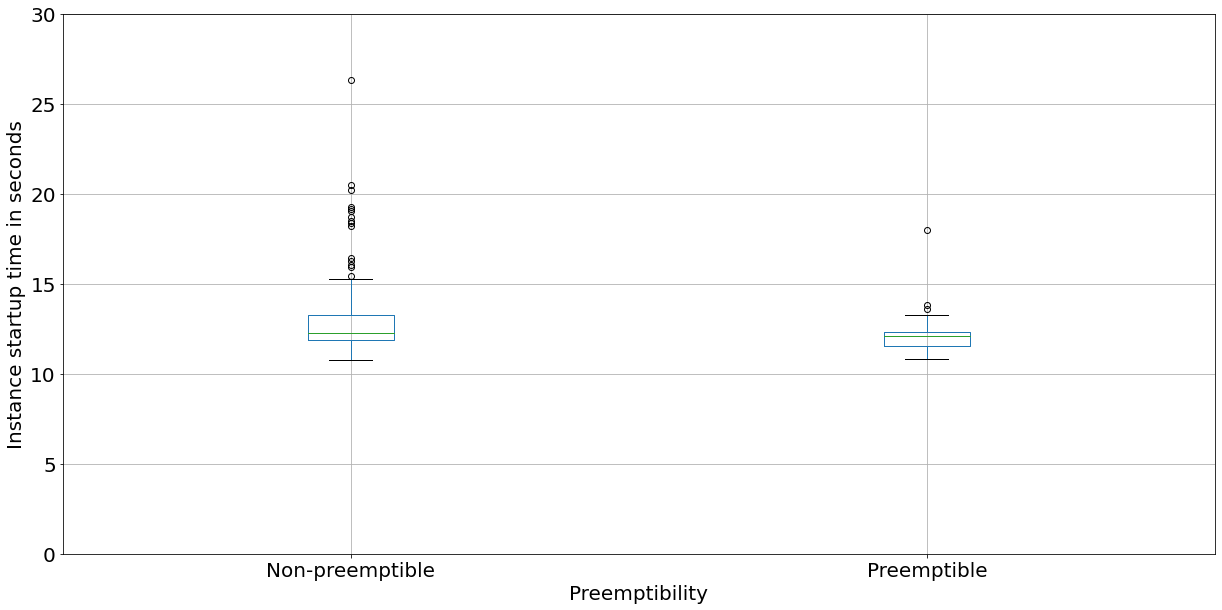}
    \caption{GCP - dependency of startup times on preemptibility}
    \label{fig:gcp_wywlaszczalnosc_czas_box}
\end{figure}

The results in~Fig.~\ref{fig:gcp_wywlaszczalnosc_czas_box} depicts that the difference is small, on average preemptibility instance are launched in $12.0$ seconds, and non-preemptible instances in $13.2$ seconds.
The preparation time is lower because such instances do not share resources with regular instances. Our research suggests no reason to avoid using a preemptible instance, at least when taking startup time into account.


\subsection{Instance deletion times}

Instance deletion times are a pertinent problem only in private clouds. In the case of most popular public clouds, there are no incurred costs in the deletion process, and thus the average user does not have any interest in the fast speed of the deletion process. On the other hand, releasing resources in private clouds is important.

\begin{figure}[!htb]
  \centering
    \includegraphics[width=0.7 \textwidth]{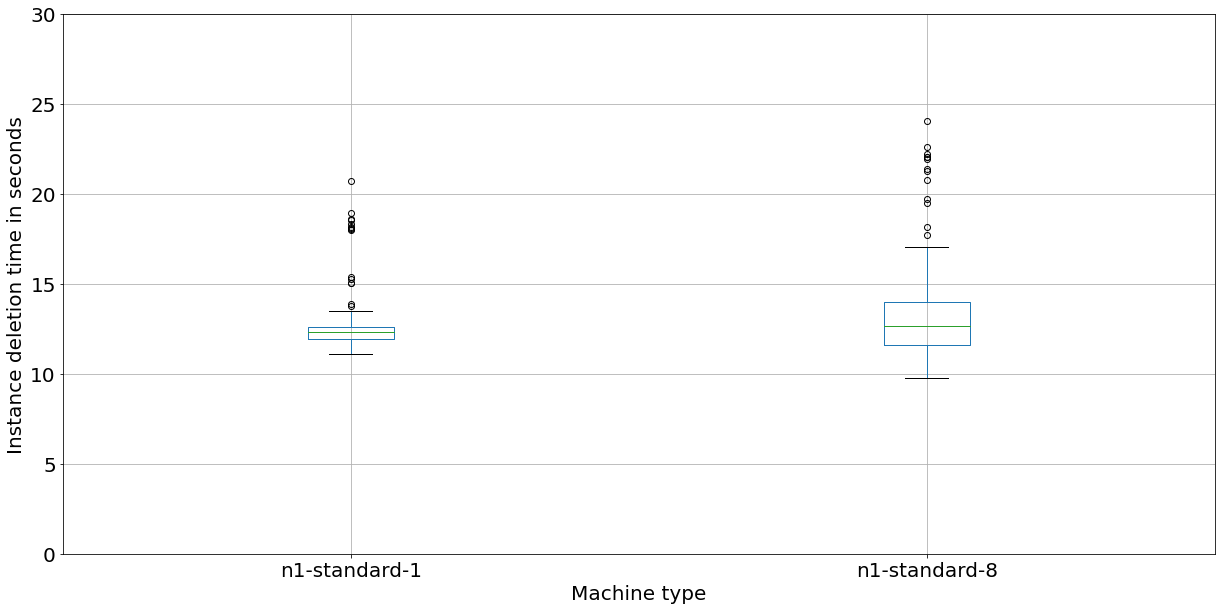}

    \includegraphics[width=0.7 \textwidth]{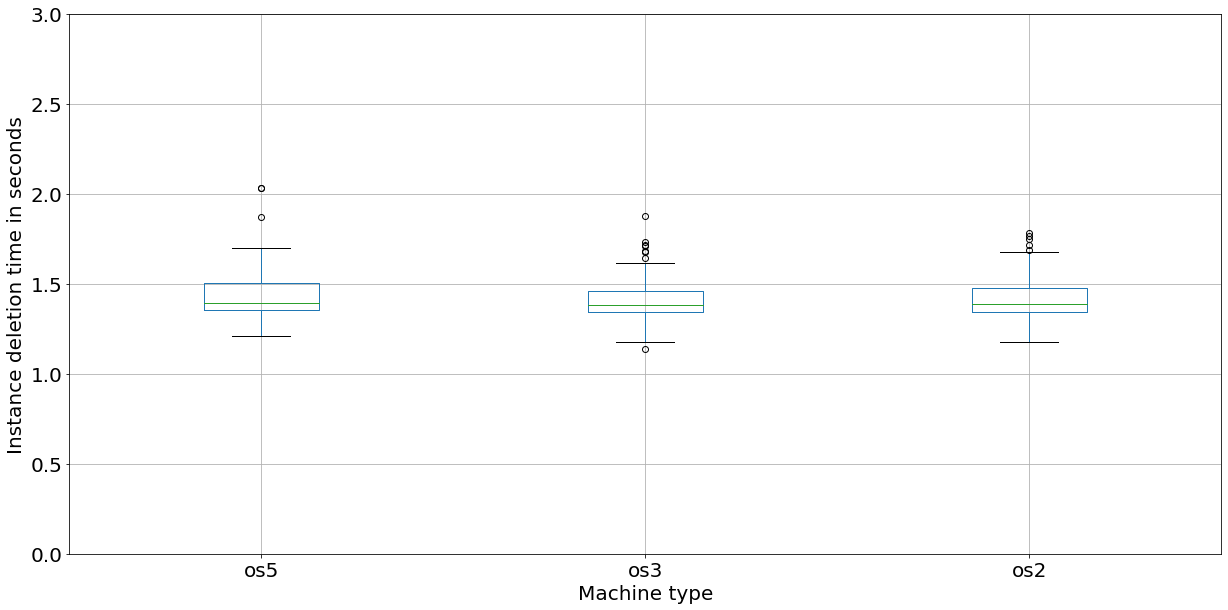}
    \caption{Dependency of deletion times, GCP on the left, OpenStack on the right.}
    \label{fig:typ_maszyny_usuwanie_box}
\end{figure}

As can be seen in~Fig~\ref{fig:typ_maszyny_usuwanie_box}, instances on GCP are usually deleted in around 120 seconds,
however, sometimes these times are closer to 20 seconds.
Such long deletion was depicted in GCP documentation~\cite{googledeleting2021}, the provider gives the user some time to undo his decision.
Due to this process, comparing the two types of machines does not allow us to reach any useful conclusions.
Deletion times in OpenStack are nearly exactly the same regardless of the system, machine type, and disk size, in our research deletion takes $1.4$ seconds on average.
This is an expected result, taking into account the results of the instance startup times analysis.

\subsection{Comparison of GCP and OpenStack in terms of instance startup times}

\begin{figure}[!htb]
  \centering
  \includegraphics[width=0.7\linewidth]{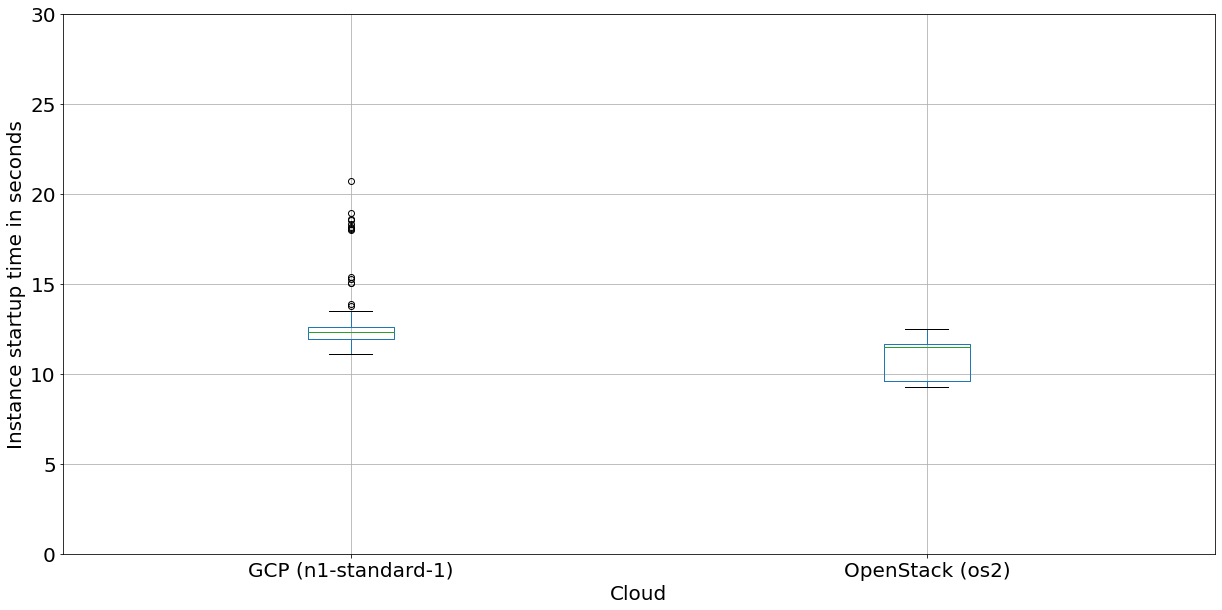}
  \caption{Instance creation times, summary on GCP and OpenStack.}
  \label{fig:gcp_openstack_porownanie_box}
\end{figure}

Fig~\ref{fig:gcp_openstack_porownanie_box} presents the speed and the changes in the speed of creating an instance on OpenStack and Google Cloud
for similar instances.
As the previous experiments showed, the machine type does not affect the startup times.

We see a significant difference of over 2 seconds, $10.8$ seconds for OpenStack, and $12.9$ seconds for GCP (on average).

\subsection{Instance performance comparison}

In this section, we compare the results obtained by running UnixBench tool~\cite{unixbench2021} 10~times on GCP and OpenStack instances. 

UnixBench is a set of benchmark algorithms that measure a broad spectrum of instance performance measurements.
Following the attached link~\cite{unixbench2021} details about each test can be found.
The UnixBench measures performance for each test (for example, in seconds for calculation, kBps for copying). Next, the measured value is compared with the BaseLine instance to produce the score's quotient (how many times the tested instance is better than the baseline instance).
The scores do not correspond directly to any physical unit.
The scores are further combined to make an overall index for the system.

The first experiment is based on the comparison of single-threaded instances,
the results were depicted in~Tab.~\ref{tab:tabela14}.

\begin{table}[!htb] \centering
\caption{Results of UnixBench tests for $1 \times$ vCPU, 1 GB RAM instances, $\mu$ is average value, $\sigma$ denotes standard deviation.}
\label{tab:tabela14}
\begin{tabular} {|l| r| r| r| r|} \hline
  Test name & \multicolumn{2}{c|}{GCP}  & \multicolumn{2}{c|}{OpenStack}  \\
  ~    & $\mu$ & \multicolumn{1}{c|}{$\sigma$}      & $\mu$ & \multicolumn{1}{c|}{$\sigma$} \\ \hline

  (a) Dhrystone 2 using register variables  & $34*10^{6}$ & $0.02\mu$ & $57*10^6$ & $0.01\mu$
  \\ \hline
  (b) Double-Precision Whetstone & $5.8*10^{3}$ & $0.02\mu$ & $9.7*10^{3}$ & $0.001\mu$
  \\ \hline
  (c) Execl Throughput & $3.2*10^{3}$ & $0.05\mu$ & $5.3*10^{3}$ & $0.01\mu$
  \\ \hline
  (d) File Copy 1024 bufsize 2000 maxblocks & $503*10^{3}$ & $0.07\mu$ & $861*10^{3}$ & $0.01\mu$
  \\ \hline
  (e) File Copy 256 bufsize 500 maxblocks & $137*10^{3}$ & $0.06\mu$ & $221*10^{3}$ & $0.01\mu$
  \\ \hline
  (f) File Copy 4096 bufsize 8000 maxblocks & $1.4*10^{6}$ & $0.12\mu$ & $2.59*10^{6}$ & $0.002\mu$
  \\ \hline
  (g) Pipe Throughput & $681*10^{3}$ & $0.06\mu$ & $1.1*10^{6}$  & $0.03\mu$
  \\ \hline
  (h) Pipe-based Context Switching &  $146*10^{3}$ & $0.05\mu$ & $248*10^{3}$ & $0.02\mu$
  \\ \hline
  (i) Process Creation & $9.7*10^{3}$ & $0.03\mu$ & $15*10^{3}$ & $0.02\mu$
  \\ \hline
  (j) Shell Scripts (1 concurrent) & $6.8*10^{3}$ & $0.04\mu$ & $11.8*10^{3}$ & $0.01\mu$
  \\ \hline
  (k) Shell Scripts (8 concurrent) & 887 & $0.04\mu$ & $1.6*10^{3}$ & $0.01\mu$
  \\ \hline
  (l) System Call Overhead & $396*10^{3}$ & $0.04\mu$ & $604*10^{4}$  & $0.004\mu$
  \\ \hline

\end{tabular}
\end{table}

The conclusions are surprising, to say the least. Not only is OpenStack more stable (1\% coefficient of variation compared to 5\% for GCP), it also offers better performance in literally any UnixBench test. As a reminder, in the experiment, we tried to eliminate the influence of virtually every factor that the user can control. OpenStack achieved an average of 61\% better performance.

There are many factors that could have influenced the result, such as hardware heterogeneity. However, it seems that, as in the article \cite{leitner2016patterns}, hardware heterogeneity does not play a big role here - it mainly affects the coefficient of variation, and the averages differ most significantly. There are many aspects the impact of which we cannot measure for legal reasons (because the public cloud provider does not provide such information) or for practical reasons (because the architecture of the private cloud is so complex that it is impossible to isolate certain factors). These include, for example: numerous layers of middleware present in IaaS systems (including the program managing the allocation of resources to individual instances), the virtualization technology used, or the noisy neighbour problem. The most significant factor that was not taken into account in the study yet is the type of processor. It is true that both types of instances had 1 virtual processor, but these processors could be based on a completely different physical infrastructure. Two subsections later we will analyse this aspect as well.

We analyse UnixBench results not only in the context of comparing the values between two instances, but also to compare parallel versions versus single-threaded instances, as depicted in~Tab.~\ref{tab:tabela15}.

\begin{table}[!htb] \centering
  \caption{Results of UnixBench tests on $8 \times$ vCPU, 16 GB RAM instances.
  Two versions of the benchmark scripts were used, benchmark using single process (sequential, seq.) and benchmark with 8 processes (concurrent, conc.).
  Test names, $\mu$ and $\sigma$ definitions are depicted in~Tab~\ref{tab:tabela14}.
}
\label{tab:tabela15}

\begin{tabular} {|l| r| r| r| r| r| r| r| r|} \hline
  Test & \multicolumn{2}{c|}{GCP seq.}  & \multicolumn{2}{c|}{GCP conc.} & \multicolumn{2}{c|}{OpenStack seq.} & \multicolumn{2}{c|}{OpenStack conc.} \\
  ~    & $\mu$ & $\sigma$      & $\mu$ & $\sigma$ & $\mu$ & $\sigma$ & $\mu$ & $\sigma$ \\ \hline
  (a) & $34*10^{6}$ & $0.01\mu$ & $187*10^{6}$ & $0.03\mu$ & $45.5*10^{6}$ & $0.01\mu$ & $255*10^{6}$ & $0.01\mu$
  \\ \hline
  (b) & $5.7*10^{3}$ & $0.001\mu$ & $41*10^{3}$ & $0.03\mu$ & $82*10^{3}$ & $0.001\mu$ & $58.5*10^{3}$ & $0.001\mu$
  \\ \hline
  (c) & $3.1*10^{3}$ & $0.02\mu$ & $14.7*10^{3}$ & $0.03\mu$ & $4.2*10^{3}$ & $0.01\mu$ & $21.2*10^{3}$ & $0.01\mu$
  \\ \hline
  (d) & $468*10^{3}$  & $0.03\mu$ & $674*10^{3}$ & $0.04\mu$ & $682*10^{3}$  & $0.01\mu$ & $1.03*10^{6}$ & $0.01\mu$
  \\ \hline
  (e) & $129*10^{3}$ & $0.03\mu$ & $183*10^{3}$ & $0.03\mu$ & $176*10^{3}$  & $0.01\mu$ & $274*10^{3}$ & $0.01\mu$
  \\ \hline
  (f) & $1.3*10^{6}$ & $0.06\mu$ & $2.1*10^{6}$ & $0.05\mu$ & $1.8*10^{6}$  & $0.01\mu$ & $3.2*10^{6}$ & $0.01\mu$
  \\ \hline
  (g) & $662*10^{3}$ & $0.04\mu$ & $3.45*10^{6}$  & $0.04\mu$ & $906*10^{3}$ & $0.004\mu$ & $4.96*10^{6}$  & $0.004\mu$
  \\ \hline
  (h) & $50*10^{3}$ & $0.07\mu$ & $626*10^{3}$ & $0.04\mu$ & $178*10^{3}$ & $0.03\mu$ & $924*10^{3}$  & $0.03\mu$
  \\ \hline
  (i) & $5.82*10^{3}$ & $0.07\mu$ & $29.6*10^{3}$ & $0.03\mu$ & $8.95*10^{3}$ & $0.02\mu$ & $39.6*10^{3}$ & $0.01\mu$
  \\ \hline
  (j) & $8.1*10^{3}$ & $0.03\mu$ & $28.8*10^{3}$ & $0.1\mu$ & $6.2*10^{3}$ & $0.05\mu$ & $4.1*10^{3}$ & $0.01\mu$
  \\ \hline
  (k) & $3*10^{3}$ & $0.02\mu$ & $4.2*10^{3}$ & $0.11\mu$ & $3.9*10^{3}$ & $0.01\mu$ & $6.3*10^{3}$ & $0.01\mu$
  \\ \hline
  (l) & $395*10^{3}$  & $0.02\mu$ & $1.95*10^{6}$  & $0.02\mu$ & $535*10^{3}$ & $0.01\mu$ & $2.86*10^{6}$ & $0.01\mu$
  \\ \hline
\end{tabular}%

\end{table}

The fully scalable algorithms should get about $8\times$ better results in 8 cores instances.
In our experiments for the file system-related tests,
we did not observe significant improvement on multi-core instances, as the tests are dependent on disk I/O performance.
Please note that \emph{Shell Scripts (8 concurrent)} runs 8 scripts in parallel. Such tests cannot run faster on multi-core instances.
The interesting tests for comparing CPU performance are \emph{Dhrystone}, \emph{Whetstone}, \emph{Execl throughput}, \emph{Pipe throughput}, and \emph{Process creation}.
We get an improvement in performance, in GCP $5.17\times$, in OpenStack $5.54\times$. It is close to the theoretical maximum, $8\times$.
It is worth mentioning that OpenStack is better at using CPU concurrency.

GCP allows the user to create instances using the processor type N1 or N2 (or others, not mentioned here).
Tab.~\ref{tab:tabela16} allow to compare GCP instances with N2 processor with OpenStack instances.
For such configuration, the OpenStack instance is only $1.11\times$ more efficient than GCP.
The processor type is important for the performance of instances on GCP, especially for concurrent algorithms.

\begin{table}[!htb] \centering
  \centering
  \caption{Results of UnixBench tests on GCP instance with $8 \times$ vCPU, 16 GB RAM and N2 processor.
  Two versions of the benchmark scripts were used (sequential, concurrent), as depicted in~Tab~\ref{tab:tabela15}.
  Test names, $\mu$ and $\sigma$ definitions are depicted in~Tab~\ref{tab:tabela14}.
}
\label{tab:tabela16}

\begin{tabular} {|l| r| r| r| r|} \hline
  Test & \multicolumn{2}{c|}{sequential}  & \multicolumn{2}{c|}{concurrent}  \\
  ~    & $\mu$ & $\sigma$      & $\mu$ & $\sigma$ \\ \hline
(a) & $42.9*10^{6}$ & $0.004\mu$ & $246*10^{6}$ & $0.002\mu$
\\ \hline
(b) & $7.3*10^{3}$ & $0.001\mu$ & $52.2*10^{3}$ & $0.001\mu$
\\ \hline
(c) & $4*10^{3}$ & $0.01\mu$ & $18.3*10^{3}$ & $0.01\mu$
\\ \hline
(d) & $779*10^{3}$ & $0.02\mu$ & $795*10^{3}$ & $0.02\mu$
\\ \hline
(e) & $212*10^{3}$ & $0.01\mu$ & $216*10^{3}$  & $0.01\mu$
\\ \hline
(f) & $2.02*10^{3}$ & $0.06\mu$ & $2.45*10^{6}$ & $0.05\mu$
\\ \hline
(g) & $1.18*10^{6}$ & $0.003\mu$ & $5.2*10^{6}$ & $0.01\mu$
\\ \hline
(h) & $63.3*10^{3}$  & $0.01\mu$ & $861*10^{3}$ & $0.02\mu$
\\ \hline
(i) & $8.6*10^{3}$ & $0.03\mu$ & $38.1*10^{3}$ & $0.01\mu$
\\ \hline
(j) & $10.7*10^{3}$ & $0.02\mu$ & $36.8*10^{3}$ & $0.01\mu$
\\ \hline
(k) & $3.82*10^{3}$ & $0.01\mu$ & $5.4*10^{3}$ & $0.02\mu$
\\ \hline
(l) & $795*10^{3}$  & $0.003\mu$  & $3.01*10^{6}$ & $0.01\mu$
\\ \hline
 \end{tabular}
\end{table}

\begin{figure}[!htb]
    \centering \includegraphics[width=1\linewidth]{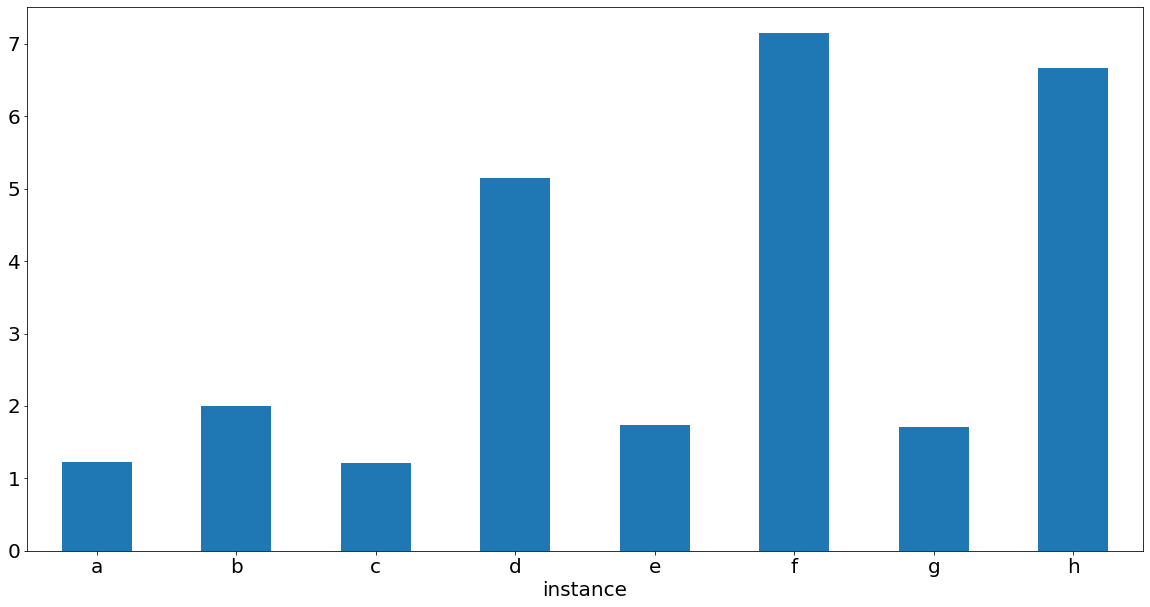}
    \caption{Benchmark comparison:
    (a): GCP - N1 instance with 1 core;
    (b): OpenStack - instance with 1 core;
    (c): GCP - N1 instance with 8 cores, non-concurrent benchmark;
    (d): GCP - N1 instance with 8 cores, concurrent benchmark;
    (e): OpenStack - instance with 8-cores, non-concurrent benchmark;
    (f): OpenStack - instance with 8-cores, concurrent benchmark;
    (g): GCP - N2 instance with 8-cores, non-concurrent benchmark;
    (h): GCP - N2 instance with 8-cores, concurrent benchmark.
    }
    \label{fig:porownanie_benchmarkow}
\end{figure}

Fig.~\ref{fig:porownanie_benchmarkow} depicts an overall comparison for the tested instances, this index is combination of all 12 tests presented in~Tab~\ref{tab:tabela14}, Tab~\ref{tab:tabela15} and Tab~\ref{tab:tabela16}. For each test, the lowest benchmark of all the tested instances (a) - (h) was found, then the value of this benchmark was used to scale the results of that test. After scaling the lowest benchmark has score$=1.0$, the results for other instances have benchmark $ \geq 1$.
Fig.~\ref{fig:porownanie_benchmarkow} presents average of scaled scores.

\section{Discussion}
\label{sec:discussion}

In our experiments, when we analysed the influence of a factor on performance, we eliminated or reduced other factors.
It allowed us to interpret results and give constructive conclusions properly. We noted in some previous works lack of this step.

It is not easy to clearly distinguish all the factors influencing the measurements in complex cloud architecture.
There were factors which could not be completely eliminated, such as technology under instances, such as the type of hypervisor used.
Moreover, some tests could have been more accurate if the data was collected throughout the years.

The benchmarks results should be treated as suggestions because every particular computational task has its specificity,
with its variety of databases transfers, calculations, data flow, etc.
Moreover, the benchmarks do not cover all aspects of a computational task, for example, UnixBench does not examine network performance.
We recommend not applying our conclusions directly but rather comparing the performance of the selected software system on different vendors.

Another aspect is studying the price-performance ratio and quality of services.
Such analyses are much easier to perform when comparing public clouds,
where conclusions about the advantage of one service over another seem to be complete only after taking the cost into account.
For example, performance varies between regions, but a typical application still needs to consider price differences.

We would also recommend analyzing clouds using more real-life workflows, as was done in the paper \cite{nawaz2016performance}, where the authors used a custom I/O intensive workflow in their analysis. While the results of such analyses are harder to compare to results obtained in other articles, they are invaluable in providing insights.

It should be noted that the most accurate comparison results could be obtained when using the same physical infrastructure. Nevertheless, we think there is some merit and need to compare those clouds from a user's perspective, treating them as black boxes.
The presented conclusions are very significant for research teams trying to decide whether to buy physical infrastructure and maintain a private cloud or use a public cloud, paying for its' use.

\section{Conclusions}
\label{sec:conclusion}

Based on the results obtained in our experiments, the instances offered by OpenStack and Google Cloud Platform have comparable performance. OpenStack achieved $5\%$ better performance in a single run situation
and $10\%$ better performance in concurrent multiple startup situations when compared to GCP. 
However, it is worth noting that the processor types of compared instances were not necessarily exactly the same, which means that the obtained results do not imply the superiority of one of those clouds.

Both discussed technologies effectively eliminated many initial differences, for example, the instance launch times between systems, machine types, times of the day.
Both platforms effectively support concurrency.

\vspace{6pt}


\section*{Availability and requirements}

Data used for this research is available on request.

\section*{Competing interests}
The~author declares that he has no competing interests.

\section*{Author contributions}

M.Ł. and R.N. identified the problem and designed the approach, M.Ł. downloaded the data, implemented the software, and performed numerical experiments. M.Ł. and R.N. interpreted the results and prepared the manuscript. R.N. provided funding. All authors have read and agreed to the published version of the manuscript.

\section*{Acknowledgements}

This research was funded by Warsaw University of Technology grant CYBERIADA-1.
We would like to thank the technical support of OpenStack at WUT.

\appendix

\section[\appendixname~\thesection]{Parameters of scripts to measure cloud performance}
\label{app:parameters}

Our environment can be used in a Unix-command-like manner. Below is the list of flags that were used to modify the parameters, enabling us to carry out our experiment.

Flags used in both clouds:
\begin{itemize}
    \item \textit{-script} - allows to specify the path to the script to be executed as soon as the instance is ready, by default \textit{None};
    \item \textit{-number} - number of instances to create, by default 1;
    \item \textit{-nodelete} - with this flag set, instances will not be deleted, they will only be created. Useful if the user does not want to delete the instances, or if he wants to delete them only after some time;
    \item \textit{-file} - name of the file where the measurements results will be saved, for GCP by default \textit{gcp\_test\_results.csv}, for OpenStack by default \textit{openstack\_test\_results.csv};
    \item \textit{-name} - the name of the instance that will be created, by default \textit{test\_instance}. If more than one instance is created, the naming will look like \textit{,,name-\#\#\#\#''}, with number of instance (0001, 0002 etc.) instead of \textit{\#\#\#\#};
    \item \textit{-iter} - allows the user to set the appropriate number of iterations of the experiment. By default, one iteration will be performed. If the flag is set \textit{nodelete}, the number of iterations, regardless of the given value, will be 1 anyway - the operation of these two flags is contrary to the logic of the program (either we perform the test cycle or we do not delete the instance, but due to the uniqueness of the names, we cannot create the same instance 100 times);
    \item \textit{-sleep} - using this flag we set the pause between experiments (in minutes). By default, this pause is set to 0, which means that subsequent experiments will be performed immediately after the end of the previous one;
    \item \textit{-image} - in both OpenStack and Google Cloud, each instance must be based on a system image. If this value is not given, the one from the configuration file will be used. In the case of Google Cloud, the user should provide a link to the image (because it can be both a public and private image) - it is a result of the fact that the format of the links accepted by the API is strictly defined, I give an example below:

    \textit{https://www.googleapis.com/compute/v1/projects/project-name/global/images/clear-linux}

    The \textit{https://www.googleapis.com/compute/v1/projects/} part is constant, then the user needs to enter a specific fragment, which can be found by going to the list of images on Google Cloud Platform and clicking the image we are interested in. In the case of OpenStack, it is enough to enter the name of one of the available images (they can be found in the Images tab in the Horizon interface).

\end{itemize}
Flags unique to GCP:
\begin{itemize}
    \item \textit{-disk} - this flag can be used to set the size of the associated disk in gigabytes, the default value is zero. The inability to set this flag in OpenStack is due to the association of the machine type with the disk (GCP does not have this association);
    \item \textit{-preemptible} - this flag allows the user to set instances as preemptible, which means that Google Compute Engine can interrupt at any time (and certainly will delete them after 24 hours). Instances of this type are not always available \cite{googlepreemptible2021}. This type of instance is not available in OpenStack;
    \item \textit{-type} - this flag is used to set the machine type. If this value is not given, the value from the configuration file will be used.
\end{itemize}

Flags unique to OpenStack:
\begin{itemize}
    \item \textit{flavor} - this flag has been extracted to conform with the OpenStack nomenclature, it allows the user to specify a system image name (and thus does not functionally differ from the "-type" flag). If a flavor name is not given, the value will be taken from the configuration file.
\end{itemize}

To connect to Google Cloud one should:
\begin{itemize}
    \item Create an account on Google Cloud.
    \item Create a service account and download the associated key as a JSON file to the selected folder.
    \item Set the variables in the configuration file according to the instructions in it.
    \item Invoke the program by running the \textit{gcp\_connector.py} file in Python with the appropriate parameters.
\end{itemize}

To connect to OpenStack one should:
\begin{itemize}
    \item Create an account in one's OpenStack implementation.
    \item Log in and download so-called ,,RC file'', tied to one's account, from the Horizon interface
    \item Execute command (and supply the OpenStack password):
    \begin{verbatim}
        source <RC file name>
    \end{verbatim}
    \item Invoke the program by running the \textit{openstack\_connector.py} file in Python with the appropriate parameters.
\end{itemize}

\bibliographystyle{abbrv}
\bibliography{bibliografia.bib}

\end{document}